\documentstyle[prl,aps]{revtex}

\newcommand{\beq}{\begin{equation}}
\newcommand{\eeq}{\end{equation}}
\newcommand{\bqn}{\begin{eqnarray}}
\newcommand{\eqn}{\end{eqnarray}}
\newcommand{\bqns}{\begin{eqnarray*}}
\newcommand{\eqns}{\end{eqnarray*}}
\newcommand{\bary}{\begin{array}}
\newcommand{\eary}{\end{array}}
\newcommand{\non}{\nonumber}
\begin{document}
\twocolumn[\hsize\textwidth\columnwidth\hsize\csname@twocolumnfalse\endcsname
\title{Reflection of matter waves by a moving wall}
\author{\small Pi-Gang Luan\\
{\footnotesize \it Institute of Electrophysics, National
Chiao-Tung University, Hsinchu, Taiwan 30043, Republic of China}}
\author{\small Yee-Mou Kao\\
{\footnotesize \it Institute of Physics, National Chiao-Tung
University, Hsinchu, Taiwan 30043, Republic of China}}
\date{\today}
\draft\maketitle

\begin{abstract}
Reflection of a normal incident matter wave by a perfectly
reflecting wall moving with a constant velocity is investigated. A
surprising phenomenon is found-that if the the wall moves faster
than the phase velocity of the incident wave, both the reflected
and incident waves propagate in the same direction. This
counter-intuitive result is an example which shows that common
sense is not always credible when one deals with quantum
problems. \vspace{5mm}
\end{abstract}]
The study of the exact solution of the time-dependent
Schr\"{o}dinger equation has drawn much attention over the past
decades\cite{Lewis,Guedes,Feng,Bauer}. Besides the mathematical
interest, the knowledge of the solution may help us to further
explore various fascinating quantum phenomena and to clarify some
subtle concepts\cite{BB,Na,Un}. In classical physics, the
knowledge of space-time transformation is essential in
understanding many interesting phenomena, such as {\it Doppler
effect}\cite{Gupta,Gleiser,Meter} and {\it optical black
holes}\cite{Leo}. However, until recently the importance of the
space-time transformation in nonrelativistic quantum systems
\cite{Green1,Green2,Vand} has not yet been emphasized or even
discussed in most popular textbooks\cite{Fey,Sakurai}. In this
report we study the Doppler effect of the matter waves echo in a
one-dimensional quantum system. We find that under appropriate
conditions this system shows surprising result. We believe the
phenomenon described below is important and deserves more
discussions and investigations.

Consider a particle of mass $m$ and momentum $p=\hbar k$ incident
from the left is reflected by a wall moving with a constant
velocity $v$. The total matter wave $\psi$ satisfies the
Schr\"{o}dinger equation
$i\hbar\partial_t\psi(x,t)=-(\hbar^2/2m)\partial^2_x\psi(x,t)$
(here $x<vt$, and $vt$ represents the position of the wall at time
$t$ ) and is the sum of the incident wave $\psi_{+}=e^{i(kx-\omega
t)}$ and reflected wave $\psi_{-}=r e^{i(k'x-\omega' t)}$: \beq
\psi(x,t)=e^{i(kx-\omega t)}+ r e^{i(k'x-\omega'
t)}\label{psitot},\eeq here $\omega=\hbar k^2/2m$ and
$\omega'=\hbar k'^2/2m$.

Suppose the moving wall is a perfectly reflecting wall. By
definition a perfectly reflecting wall is the boundary separates
the regions of potential $V(x)=0$ and $V(x)=\infty$ where wave
function vanishes. Since the wall moves uniformly with velocity
$v$ we have the boundary condition \beq \psi(vt,t)=0,\label{bc}
\eeq which leads to the obvious solution \beq
r=-1,\;\;\;\;\;k'=-k+\frac{2mv}{\hbar}\label{rk}, \eeq and hence
the phase velocity $v_p$ of the reflected wave $\psi_{-}$ is given
by \beq v_p=\frac{\hbar}{2m}(-k+\frac{2mv}{\hbar})=v-\frac{\hbar
k}{2m}. \eeq

Up to now everything seems simple and reasonable. For a wall moves
toward the left or moves slowly enough toward the right we have
$k'<0$ and the reflected wave propagates toward the opposite
direction of the incident wave, which is consistent with our naive
intuition. However, what will happen if the wall moves fast enough
so that $2mv>\hbar k$? Surprisingly, in this situation $k'>0$ and
hence the incident and reflected waves propagate toward the same
direction! Furthermore, if we increase the velocity of the wall
such that $v>\hbar k/m$ then the reflected wave will propagate
faster than the incident wave. One might feel uncomfortable and
doubt if this counter-intuitive phenomenon will actually happen.
However, since the above derivation based merely on the
Schr\"{o}dinger equation itself and the simple boundary condition
(\ref{bc}), the result must be true under the given conditions.

Substitute (\ref{rk}) into (\ref{psitot}), and define \beq
\bar{x}=x-vt,\;\;\;\;\;\bar{k}=k-\frac{mv}{\hbar}, \eeq we have
\beq \psi(x,t)
=\exp{\left[i\left(\frac{mv}{\hbar}x-\frac{mv^2}{2\hbar}t\right)\right]}\varphi(\bar{x},t),
\label{Gali} \eeq where \beq
\varphi(\bar{x},t)=2i\sin(\bar{k}\bar{x})
\exp{\left(-i\frac{\hbar\bar{k}^2}{2m}t\right)}. \eeq Note that
Eq.(\ref{Gali}) is nothing but the Galilean Transformation from
the original reference frame to another one moving with velocity
$v$ with respect to the first, and $\varphi(\bar{x},t)$ is the
wave function in that system. This observation leads us to a
different understanding of Eq.~(\ref{rk}). Denote the original
reference frame as $S$ and the second one as $S'$. In $S$ the
incident wave $\psi_{+}$ has wave number $k$ and the reflected
wave $\psi_{-}$ has wave number $k'=-k+2mv/\hbar$, as discussed
before. On the other hand, in $S'$ there are two different
possibilities. If $\bar{k}=k-mv/\hbar>0$ we see a incident wave
with wave number $\bar{k}$ and a reflected wave with wave number
$-\bar{k}$; whereas if  $\bar{k}=k-mv/\hbar<0$ we see a incident
wave with wave number $-\bar{k}$ and a reflected wave with wave
number $\bar{k}$. Thus in $S'$ which wave component is defined as
the incident wave is determined by the sign of the wave number
$\bar{k}$. Consequently, Eq.~(\ref{rk}) is simply established by
transforming the wave function in $S'$ back to $S$.

From these results we have the unnormalized probability density
\beq |\psi|^2
=4\sin^2\left[\left(k-\frac{mv}{\hbar}\right)\left(x-vt\right)\right]
\eeq and the probability current \bqn
J&=&(\hbar/2mi)(\psi^*\partial_x\psi-\psi\partial_x\psi^*)\non\\
&=&4v\sin^2\left[\left(k-\frac{mv}{\hbar}\right)\left(x-vt\right)\right].
\eqn

Now, a ``drift velocity" $v_d$ can be defined as the ratio
$J/|\psi|^2$, and this definition gives us the reasonable result
\beq v_d=v. \eeq This means that the pattern of the particle
probability density $|\psi|^2$ behind the wall is dragged by the
wall and moves uniformly with velocity $v$.

In conclusion, we have shown that under appropriate conditions
even the simplest one-dimensional quantum scattering shows
unexpected results. We believe the phenomenon described in this
report is important and hope it can stimulate more related
investigations.

This work received support from National Science Council.

\end{document}